\documentclass[ reprint, amsmath,amssymb, aps,nofootinbib]{revtex4-1}
\usepackage{graphicx}
\usepackage{bm}
\graphicspath{ {./images/} }

\usepackage[dvipsnames]{xcolor}
\usepackage{footnote}
\usepackage{enumitem}
\usepackage{qcircuit}
\usepackage{float}
\usepackage{mathtools}
\usepackage{amsmath}
\usepackage{subcaption}
\usepackage[colorlinks]{hyperref}

\usepackage{verbatim}

\usepackage{caption}
\DeclarePairedDelimiter{\ceil}{\lceil}{\rceil}

\begin{document}

\title{ Implementation of a general single-qubit positive operator-valued measure on a
circuit-based quantum computer}

\author{Yordan S. Yordanov}
\author{Crispin H. W. Barnes}
\affiliation{Cavendish Laboratory, Department of Physics, University of Cambridge, Cambridge CB3 0HE, United Kingdom}

\date{\today}

\begin{abstract}
We derive a deterministic protocol to implement a general single-qubit POVM on near-term circuit-based quantum computers. The protocol has a modular structure, such that an $n$-element POVM is  implemented as a sequence of $(n-1)$ circuit modules.  Each module performs a 2-element POVM.
Two variations of the protocol are suggested, one optimal in terms of number of ancilla qubits, the other optimal in terms of number of qubit gate operations and quantum circuit depth.
We use the protocol to implement $2$- and $3$-element POVMs on two publicly available quantum computing devices.
The results we obtain suggest that implementing non-trivial POVMs could be within the reach of the current noisy quantum computing devices.

\end{abstract}

\maketitle


\section{ Introduction}\label{Intro}
In quantum mechanics positive operator-valued measures (POVMs) describe the most general form of quantum measurement. They are able to distinguish probabilistically  between non-orthogonal quantum states \cite{nonOrtho} and can therefore be used to perform optimal state discrimination \cite{discrimination1, POVM2_1} and efficient quantum tomography \cite{tomography1, tomography2}. In quantum communication and cryptography\cite{comm}, they are used to enable secure device independent communication \cite{comm1}, or, on the contrary, compromise quantum key distribution protocols by minimizing the damage done by an eavesdropper to a quantum channel \cite{QKD1,QKD2}.

POVMs can be implemented experimentally in both bosonic \cite{AnP, AnP2, POVM6, POVM7} and fermionic quantum systems \cite{POVM4}.  However, typically, the hardware for these implementations needs to be specifically tailored to the measurement. To realize an arbitrary POVM as part of quantum communication scheme or on a quantum computer, where the hardware design allows only orthogonal projective measurements in the qubit basis, it is necessary to simulate the action of the POVM using quantum-gate operations. For example, in reference \cite{POVM3} a quantum Fourier transform is used to implement a restricted class of projective POVMs. In references \cite{POVM1, POVM8} a probabilistic method, based on classical randomness and post-selection, is proposed to implement projective POVMs. A deterministic method to perform a general POVM can be implemented using Neumark's dilation theorem\citep{Naimark0,Naimark0.1}, which states that a POVM of $n$ elements can be performed as a projective measurement in a $n$-dimensional space. In reference \cite{POVM5} it is shown that this method can be realized in a duality quantum computer.

In this work we construct a protocol for a general single-qubit POVM on a circuit-based quantum computer, using Neumark's theorem.  The protocol has a modular structure such that a quantum circuit for a $n$-element POVM is constructed as a sequence of $(n-1)$ $2$-element POVM circuit modules, in a similar manner to reference \cite{AnP}.  This structure allows for a straightforward construction of quantum circuits, using an optimal number of ancilla qubits and quantum gates.
The complexity of the protocol, in terms of number of quantum gates, is $O(n^2)$ using $\ceil{\log_2 n}$ ancilla qubits, and can be reduced to $O(n \log n)$ at a cost of $(\ceil{\log_2n}-1)$ additional ancilla qubits. The corresponding circuit depths are $O(n^2)$ and $O(n)$ respectively. 
We use the protocol to implement $2$- and $3$-element POVMs on two public quantum computing devices; IBMQX2 and Aspen4. We measure the output fidelities and compare the performances of the two devices.

In sec. \ref{Method} we present our protocol. We describe explicitly how to construct a quantum-gate circuit for a $2$-element POVM, and demonstrate how it can be extended to a $n$-element POVM.
In sec. \ref{real_devices} we present the results from the POVM implementations on the two quantum devices. We present our concluding remarks in sec. \ref{discussion}.

\section{ POVM Protocol}\label{Method}

\textbf{Preliminaries: }
An $n$-element POVM is defined as a set of $n$ positive operators $\{\hat{E}_i\}$ that satisfy the completeness relation $ \sum_{i=1}^{n} \hat{E}_i = \hat{I}$, where $\hat{E}_i = \hat{M}_i^\dagger \hat{M}_i$ and the $\{\hat{M}_i\}$ are measurement operators. Performing a POVM on a system in initial state $|\psi_0\rangle$ results in wave function reduction to one of $n$ possible measurement outcomes $|\psi_0\rangle \rightarrow |\psi_i\rangle =\frac{\hat{M}_i|\psi_0\rangle}{\sqrt{\langle\psi_0|\hat{M}_i^{\dagger}  \hat{M}_i|\psi_0\rangle}}$, with probability $p_i = \langle\psi_0|\hat{M}_i^{\dagger}  \hat{M}_i|\psi_0\rangle$. Using Neumark's theorem, a $n$-element POVM on a target system A, can be performed by introducing an ancilla system B, with Hilbert space spanned by $n$ orthonormal basis states $|i^{(B)}\rangle$ that are in one-to-one correspondence with the POVM measurement outcomes. A unitary operation $\hat{U}_{AB}$ is applied to the joint state of the two systems, such that
\begin{equation}\label{eq.Uab}
\hat{U}_{AB}|\psi_0^{(A)}\rangle|0^{(B)}\rangle = \sum_{i=1}^n\big[\hat{M}_{i}|\psi_0^{(A)}\rangle\big]|i^{(B)}\rangle.
\end{equation}
By performing a projective measurement on system B, system A collapses to one of the $n$ states $\hat{M}_i|\psi^{(A)}_0\rangle$ that correspond to the outcomes of the POVM. For more details on POVM implementation refer to \cite{POVM, SuperOps}.

\textbf{Protocol outline:  }
Based on the method, described above, we implement a $n$-element POVM on a target system consisting of a single qubit, using an ancilla system of $\ceil{\log_2n}$ qubits. To implement $\hat{U}_{AB}$, we divide it into a sequence of $(n-1)$ quantum gate circuits, which we call modules. Each of these modules, except the first, performs a $2$-element POVM on one of the outcomes of the preceding module, and entangles the additionally produced outcome to a new state of the ancilla system.

\begin{figure}[H]
\centering

    \includegraphics[width=0.48\textwidth]{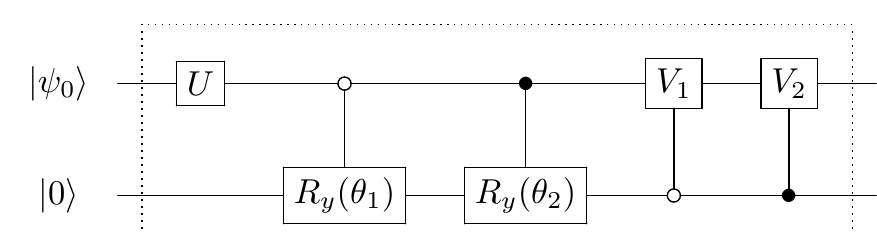}
    \caption{A quantum circuit for a general single-qubit $2$-element POVM. The top qubit acts as the target, and the bottom as the ancilla. The output state of the circuit is given by eq. (\ref{eq:2POVM_res}). $\hat{R}_y(\theta)$ denotes a controlled single-qubit $y$-rotation by angle $\theta$. $\hat{U}$, $\hat{V}_1$ and $\hat{V}_2$ denote general single-qubit unitary operations. $\hat{V}_1$ and $\hat{V}_2$ are controlled operations, and each of them can be implemented as a combination of controlled $z-$ and $y-$rotations. The circuit contains up to $12$ CNOTs and $14$ single-qubit rotations.\\
    $\_\_\_\_\_\_\_\_\_\_\_\_\_\_\_\_\_\_\_\_\_\_\_\_\_\_\_\_\_\_\_\_\_\_\_\_\_\_\_\_\_\_\_\_\_\_\_\_\_\_\_\_\_\_\_\_\_\_\_\_\_\_\_\_\_\_\_$}
    \label{fig:2POVM}
\end{figure}


\textbf{2-element POVM module: }
To construct a quantum circuit performing a $2$-element POVM, we need a single ancilla qubit.
We assume the target qubit starts in an arbitrary state $|\psi_0 \rangle = a|0\rangle + b|1\rangle$. Then the initial state of the system, target plus ancilla, is $|\Psi_0 \rangle = |\psi_0 \rangle|0\rangle $.
To perform a $2$-element POVM we want to transform the system to a state
\begin{equation}\label{eq:POVM2_out.0}
|\Psi_f\rangle = \big(\hat{M}_1|\psi_0\rangle)|o_1\rangle + \big(\hat{M}_2|\psi_0\rangle)|o_2\rangle,
\end{equation}
where $\hat{M}_1$ and $\hat{M}_2$ are the two measurement operators and $|o_1\rangle$ and $|o_2\rangle$ are two orthogonal states of the ancilla.
First a unitary gate $\hat{U}$ (not to be confused with $U_{AB}$) is performed on the initial state of the target qubit:
\begin{equation}
|\Psi_0\rangle \rightarrow \big(\hat{U}|\psi_0\rangle\big)|0\rangle = \big(a'|0\rangle +b'|1\rangle\big)|0\rangle.
\end{equation}
Then, two controlled $y$-rotations are performed, acting on the ancilla qubit and controlled by the target qubit.
The rotations are given by angles $\theta_1$ and $\theta_2$, and controlled by the target qubit in states $|0\rangle$ or $|1\rangle$ respectively. 
\begin{equation}\label{eq:theta12}
|\Psi \rangle \rightarrow  a' |0\rangle  (\cos\theta_1 |0\rangle + \sin\theta_1|1\rangle)+  b'|1\rangle(\cos\theta_2 |0\rangle + \sin\theta_2|1\rangle).
\end{equation}
Rearranging terms, the state above can be written as
\begin{equation}\label{eq:D1D2}
|\Psi \rangle =\big(\hat{D}_1 \hat{U}|\psi_0\rangle\big)|0\rangle + \big(\hat{D}_2 \hat{U}|\psi_0\rangle\big)|1\rangle,
\end{equation}
where $\hat{D}_1 = \cos \theta_1 |0 \rangle \langle 0 | + \cos \theta_2 |1 \rangle \langle 1 |$ and
$\hat{D}_2 = \sin \theta_1 |0 \rangle \langle 0 | + \sin \theta_2 |1 \rangle \langle 1 |$. This result
corresponds to performing a $2$-element POVM specified by arbitrary operators $\hat{E}_1$ and $\hat{E}_2$.
However to fully specify the measurement operators $\hat{M}_1$ and $\hat{M}_2$, we need to perform unitary operations $\hat{V}_1$ and $\hat{V}_2$ on the terms in the target qubit state, corresponding to the two outcomes of the POVM. This can be done by two single-qubit unitary gates acting on the target qubit, and controlled by the ancilla states corresponding to the two POVM outcomes, $|0\rangle$ and $|1\rangle$ respectively. This results in a final state
\begin{equation}\label{eq:2POVM_res}
|\Psi \rangle \rightarrow |\Psi_f\rangle= \Big(\hat{V}_1\hat{D}_1\hat{U}|\psi_0\rangle\Big)|0\rangle + \Big(\hat{V}_2\hat{D}_2\hat{U} |\psi_0\rangle\Big)|1\rangle,
\end{equation}
with $\hat{V}_1 \hat{D}_1 \hat{U} = \hat{M}_1$ and $ \hat{V}_1 \hat{D}_2 \hat{U} = \hat{M}_2 $.  Since $\hat{U}$, $\hat{V}_1$ and $\hat{V}_2$ are unitaries, and $\hat{D}_1\hat{D}_1^{\dagger} + \hat{D}_2\hat{D}_2^{\dagger} =I $, it is straightforward to check that $\hat{M}_1$ and $\hat{M}_2$ satisfy the completeness relation.
Furthermore, the expressions for the two measurement operators are in most general form, since they correspond to singular value decompositions.
Therefore eq. (\ref{eq:2POVM_res}) corresponds to the outcomes of a general $2$-element POVM.
Figure \ref{fig:2POVM} illustrates the complete circuit for the $2$-element POVM module.

\textbf{Generalization to n-element POVM }
A $n$-element POVM can be performed  sequentially by $(n-1)$ POVM modules, that share an ancilla register of $\ceil{\log_2n}$ qubits. The $i^{th}$ module in the sequence will be characterized by rotation angles $\theta^{(i)}_1$ and $\theta^{(i)}_2$, unitary operations $\hat{V}_1^{(i)}$ and $\hat{V}^{(i)}_2$, and two POVM outcomes with corresponding orthogonal ancilla register states $|o^{(i)}_1\rangle$ and $|o^{(i)}_1\rangle$.
The first module is additionally characterized by the unitary $\hat{U}$ acting on the target qubit, as shown above.
Each of the modules, except the first one, performs a $2$-element POVM on the second outcome of the preceding module, so that the term in the target qubit state, corresponding to this outcome, is evolved in a similar way as for the case of the $2$-element POVM.
The output state of the sequence of modules can be written as
\begin{equation}\label{eq:nPOVM_out}
|\Psi\rangle=\sum^{n-1}_{i=1} \Big(\hat{M}_i|\psi_0\rangle\Big)|o^{(i)}_1\rangle + \Big(\hat{M}_n|\psi_0\rangle\Big)|o^{(n-1)}_2\rangle
\end{equation}

with the measurement operators $\hat{M}_i$ given by
\begin{equation}\label{eq:Mi}
\hat{M}_{i} = \begin{cases} \hat{M}_1 = \hat{V}_1^{(1)} \hat{D}_1^{(1)} \text{ , for i =1}\\ \\
\hat{V}_1^{(i)} \hat{D}_1^{(i)} \prod_{j=1}^{i-1} \Big( \hat{V}_2^{(j)} \hat{D}_2^{(j)} \Big)  \hat{U} \text{ , for } 1<i<n \\ \\
\prod_{j=1}^{n-1} \Big( \hat{V}_2^{(j)} \hat{D}_2^{(j)}\Big)  \hat{U}  \text{ , for } i=n,

\end{cases}
\end{equation}
where $\hat{D}_1^{(i)} = \cos \theta_1^{(i)} |0\rangle \langle 0| + \cos \theta_2^{(i)} |1\rangle \langle 1|$ and $\hat{D}_2^{(i)} = \sin \theta_1^{(i)} |0\rangle \langle 0| + \sin \theta_2^{(i)} |1\rangle \langle 1|$.
These measurement operators satisfy the completeness relation, and also represent singular value decompositions as in the case of the $2$-element POVM. Therefore eq. (\ref{eq:nPOVM_out}) describes the outcomes of a general single-qubit $n$-element POVM.
Appendix \ref{app:nPOVM_steps} presents an explicit procedure for the construction of a quantum circuit for the $i^{th}$ module. This procedure can be used iteratively to construct the whole $n$-element POVM. With a few additional operations the ancilla states $|0^{(j)}_{1/2}\rangle$ can be chosen so that the $i^{th}$ POVM outcome corresponds to  the ancilla state with a binary value $(i-1)$.
The quantum circuit for a POVM module sequence is illustrated in Fig. \ref{fig:nPOVM_new} in Appendix \ref{app:nPOVM_steps}.

\textbf{Complexity and circuit depth: }
In app. \ref{app:complexity} we  show that the complexity, in terms of number of quantum gates, of the $i^{th}$ POVM module, is $O(i)$. Summing over all, modules the complexity for an $n$-element POVM is $\sum_{i=1}^{n-1} O(i)= O(n^2)$. The depth of the quantum circuit, in terms of $CNOTs$, scales quadratically with $n$ also.
Alternatively we can use $(\ceil{\log_2 n}-1)$ additional ancilla qubits  to reduce the complexity of the $i^{th}$ module to $O(\log i)$. This results in overall complexity of $O(n \log n)$ for a $n$-element POVM. In this case the circuit depth for the $i^{th}$ module is constant (at most $18$ CNOTs), hence the depth for a $n$-element POVM becomes linear in $n$.\\
In the implementation of the two POVM examples, in section \ref{real_devices}, we use the quadratic method however (that requires fewer ancilla qubits), since for $n=2$ and $n=3$, both methods use the same number of quantum gates and have equal maximum circuit depths.

\textbf{Extension to $N$-qubit POVMs: }
The modular structure of this protocol can be extended to the case of a POVM on a $d$-level system by modifying the circuit of the POVM module.
In the case of the single-qubit target system, we performed the two rotations $\theta_1$ and $\theta_2$ on the ancilla qubit  (eq. (\ref{eq:theta12})), controlled by the two states of the target qubit.
In the case of a $d$-level target system, we will have to perform $d$ rotations - specified by angles $\{\theta_{i \in [1,d]}\}$ - and controlled by the $d$ different states of the target system.
The output state of the $2$-element POVM module is, therefore, given again by eq. (\ref{eq:2POVM_res}), where this time $\hat{U}$, $\hat{V}_1$ and $\hat{V}_2$ are $d$-dimensional unitary operations,
$\hat{D}_1 = \sum_{i=0}^{d-1} \cos \theta_i |i\rangle\langle i|$ and $\hat{D}_2 = \sum_{i=0}^{d-1} \sin \theta_i |i\rangle\langle i|$.
However implementing any of $\hat{U}$, $\hat{V}_1$ and $\hat{V}_2$ now involve the generic problem of performing a general unitary operation on a multi-qubit system. Therefore, the modular structure does simplify, but does not fully solve the problem of implementing a general multi-qubit POVM.

\section{Implementation on quantum computing devices}\label{real_devices}

Using our protocol we implement a $2$- and a $3$-element POVMs on two public quantum computing devices; IBM's $5$-qubit IBMQX2 \cite{IBM}, and Rigetti's  $16$-qubit Aspen4 \cite{rigetti}. These devices are capable of performing universal operations \cite{DiVin} on their qubit registers. However they have high noise levels and imperfect qubit control, and hence are reffered to as noisy intermediate scale quantum (NISQ) devices \cite{NISQ1}.

\textbf{2-element POVM: }
First we consider an example of a $2$-element POVM that exhibits an output state with clear symmetry in terms of its outcomes. We choose two equal measurement operators, defined by $\theta_1=\theta_2=\frac{\pi}{4}$, $
\hat{V}_1 =
\hat{V}_2 =
\hat{I}$, $
\hat{U}_1 = \frac{1}{2} \bigl( \begin{smallmatrix}1 & 0\\ 0 & \sqrt{3}\end{smallmatrix}\bigr)$, and an initial target qubit state  $|\psi_0\rangle=|0\rangle$. Note that the resulting measurement operators $
\hat{M}_1=
\hat{M}_2= \frac{1}{2} \bigl( \begin{smallmatrix}1 & 0\\ 0 & \sqrt{3}\end{smallmatrix}\bigr)$ are not projective.
From eq. (\ref{eq:2POVM_res}) the expected output state is
\begin{equation}\label{eq:2POVM_example}
|\Psi\rangle = \frac{\big(|0\rangle+\sqrt{3}|1\rangle\big)|0\rangle + \big(|0\rangle+\sqrt{3}|1\rangle\big)|1\rangle}{2\sqrt{2}}.
\end{equation}

\begin{figure}[H]
    \centering
    \includegraphics[width=0.52\textwidth]{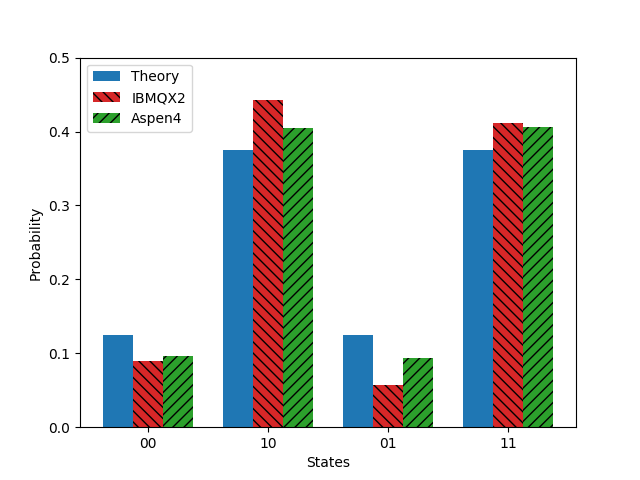}
    \caption{Measurement probabilities for a 2-element POVM, described by $
\hat{M}_1=
\hat{M}_2= \frac{1}{2} \bigl( \begin{smallmatrix}1 & 0\\ 0 & \sqrt{3}\end{smallmatrix}\bigr)$, on a qubit in initial state $|0\rangle$.
The ibmqx2 probabilities are obtained from $8192$ runs of the circuit, and the Aspen4 probabilities - from $10^4$ runs. The expected probability values from eq. (\ref{eq:2POVM_example}) are included for reference.\\
    $\_\_\_\_\_\_\_\_\_\_\_\_\_\_\_\_\_\_\_\_\_\_\_\_\_\_\_\_\_\_\_\_\_\_\_\_\_\_\_\_\_\_\_\_\_\_\_\_\_\_\_\_\_\_\_\_\_\_\_\_\_\_\_\_\_\_\_$}
    \label{fig:2POVM_all}
\end{figure}

Figure~\ref{fig:2POVM_all} presents the results for the $2$-element POVM, from the two quantum devices.
The Aspen4 output has a fidelity of $99.5\%$\footnote{The fidelity values are calculated as the overlap $F=|\langle \psi | \phi \rangle|^2$, of two pure states, instead of as $F=\langle \phi|\sigma_{\psi}|\phi\rangle$, where $\sigma_{\psi}$ is the generally mixed state produced by a real device.}. Although the outcome state of the target qubit is not obtained exactly, the expected symmetry between the states corresponding to the two POVM outcomes is obtained.
The output from the IBMQX2 is less accurate, with fidelity of $98.0\%$, exhibiting asymmetry in the measurement probabilities for the values of the ancilla qubit corresponding to the two POVM outcomes.
A possible reason for this asymmetry is the fact that IBMQX2 have different CNOT gate error rates depending on which qubit is the control or the target (see \cite{IBM} for device characterization).

\textbf{3-element POVM: }
The second example we implement is a $3$-element POVM defined by measurement operators, that project on three states separated by $\frac{2 \pi}{3} rad$ in the $x-z$ plane of the Bloch sphere:
\begin{equation}\label{eq:M31}
\hat{M}_1=\sqrt{\frac{2}{3}}|0\rangle\langle 0|,
\end{equation}
\begin{equation}\label{eq:M32}
\hat{M}_2=\frac{1}{\sqrt{6}}\frac{|0\rangle + \sqrt{3}|1\rangle}{2}\frac{\langle0| + \sqrt{3}\langle1|}{2},
\end{equation}
\begin{equation}\label{eq:M33}
\hat{M}_3=\frac{1}{\sqrt{6}}\frac{|0\rangle - \sqrt{3}|1\rangle}{2}\frac{\langle0| - \sqrt{3}\langle1|}{2}.
\end{equation}
This POVM is a classic example, often considered in literature, which can be used to distinguish between two non-orthogonal states (for example between $ |1\rangle$ and $\frac{\sqrt{3}|0\rangle+|1\rangle}{2} $). It is implemented using two POVM modules defined by; $\theta_1^{(1)} = \cos^{-1}\Big(\sqrt{\frac{2}{3}}\Big)$, $\theta^{(1)}_2 =\frac{\pi}{2}$, $ \theta_1^{(2)}=0, \theta_2^{(2)} =\frac{\pi}{2}$, $
\hat{U}=
\hat{I}$, $
\hat{V}_2^{(1)}= \frac{1}{\sqrt{2}}\bigl( \begin{smallmatrix}1 & 1\\ -1 & 1\end{smallmatrix}\bigr)$, $
\hat{V}_1^{(1)} =
\hat{I}$, $
\hat{V}_1^{(2)}= \frac{1}{2}\bigl( \begin{smallmatrix} 1 & -\sqrt{3}\\ \sqrt{3} & 1\end{smallmatrix}\bigr)$ and $\hat{V}_2^{(2)} =- \frac{1}{2}\bigl( \begin{smallmatrix}\sqrt{3} & -1\\ 1 & \sqrt{3}\end{smallmatrix}\bigr)$. In app. \ref{app:3POVM} we outline explicitly the steps to construct a quantum circuit for the second POVM module.
Substituting eqs. (\ref{eq:M31}), (\ref{eq:M32}) and (\ref{eq:M33}) in eq. (\ref{eq:nPOVM_out}), for an initial target qubit in a state $|\psi_0\rangle=|0\rangle$, the expected output state is
\begin{multline}\label{eq:3POVM_example}
\Psi =  \sqrt{\frac{3}{2}}|0\rangle|00\rangle + \frac{|0\rangle +  \sqrt{3}|1\rangle}{2\sqrt{6}}|10\rangle + \frac{|0\rangle - \sqrt{3}|1\rangle}{2\sqrt{6}}|01\rangle.
\end{multline}

\begin{figure}[H]
    \centering
    \includegraphics[width=0.52\textwidth]{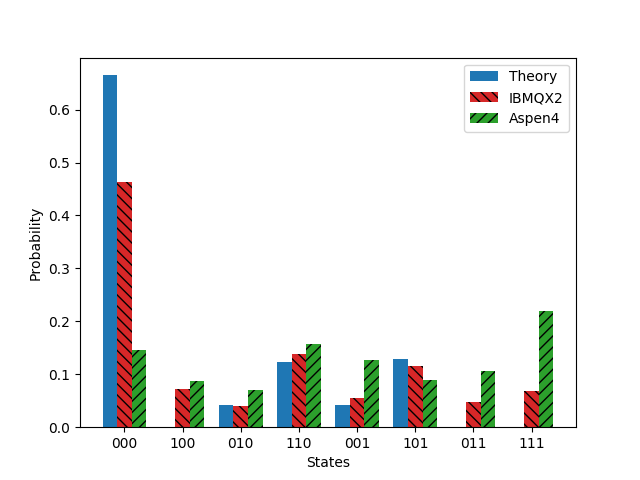}
    \caption{ Measurement probabilities for a $3$-element POVM, described by measurement operators given by eqs. (\ref{eq:M31}), (\ref{eq:M32}) and (\ref{eq:M33}), on a qubit in initial state $|0\rangle$. The IBMQX2 probabilities are obtained from $8192$ runs of the circuit, and the Aspen4 probabilities - from $10^4$. The expected probability values from eq. (\ref{eq:3POVM_example}) are included for reference. \\
    $\_\_\_\_\_\_\_\_\_\_\_\_\_\_\_\_\_\_\_\_\_\_\_\_\_\_\_\_\_\_\_\_\_\_\_\_\_\_\_\_\_\_\_\_\_\_\_\_\_\_\_\_\_\_\_\_\_\_\_\_\_\_\_\_\_\_\_$}
    \label{fig:3POVM_all}
\end{figure}

Figure \ref{fig:3POVM_all} shows the results for the $3$-element POVM, obtained from the two quantum devices.
In this case it is evident that the results from both devices suffer from significantly higher decoherence than in the case of the $2$-element POVM. The IBMQX2 performs better this time, obtaining an output state with fidelity $80.2\%$. It produces close to the expected values for the measurement probabilities of the $|010\rangle$, $|110\rangle$, $|001\rangle$ and $|101\rangle$ states. However the state $|000\rangle$ seems to have decayed to the states with zero-expected probability, $\left| 100 \right\rangle$, $\left| 011 \right\rangle$, and $\left| 111 \right\rangle$.
The output from the Aspen4 has fidelity of $46.6\%$ and demonstrates little correlation with the expected output.
The reason for these significantly worse results in the case of the $3$-element POVM is the depth and complexity of the quantum circuit. For comparison the $2$-element POVM circuit has $6$ CNOTs, resulting in a depth of $6$ also, while the $3$-element POVM circuit has $30$ CNOTs, with a maximum depth of $16$ for the target qubit.

\section{Conclusion}\label{discussion}

In this paper we presented a deterministic protocol that enables a general POVM to be performed on a qubit in a circuit-based quantum computer, using a conventional set of single and two-qubit quantum gates. We show that the same protocol can be modified so that it can be applied to several qubits. We implement the POVM as a projective measurement, using Neumark's theorem, on an ancilla register of qubits. The protocol therefore does not measure the target qubit and hence can be used as a subroutine in a larger protocol.

We use the protocol to implement a $2$- and  a $3$-element POVMs on two quantum computing devices; IBM's IBMQX2, and Rigetti's Apsen4.
In the case of the $2$-element POVM, both devices produce high fidelity results, with the Aspen4 being more accurate and consistent than the IBMQX2.
For the $3$-element POVM, the results from both devices evidently suffer from strong decoherence. Nevertheless, the results of IBMQX2, demonstrate good correlation with the expected output and fidelity of $\sim 80\%$.  This result suggests that there is reason to be optimistic that given the regular upgrades of these devices, it might soon be possible to perform these measurements with high fidelity.  This will open the way to their use in a wide variety of applications including quantum tomography and quantum cryptography.

\begin{acknowledgements}
We acknowledge financial support from the Hitachi ICASE project, and the Engineering and Physics Research Council (EPSRC). We are grateful to IBM and Rigetti for the opportunity to use their cloud-based quantum computing services. We also would like to thank A. Andreev, A. Lasek, D. Arvidsson-Shukur, H. Lepage, J. Drori and N. Devlin for useful discussions.
\end{acknowledgements}

\bibliographystyle{unsrt}
\bibliography{references}

\onecolumngrid

\appendix
\newpage

\section{Constructing the $i^{th}$ module of an $n$-element POVM}\label{app:nPOVM_steps}

Here we describe the explicit steps to construct the $i^{th}$ module of a $n$-element POVM.
The key is to entangle the two POVM outcomes of the module with suitable computational states of the ancilla register, so that one can perform the same operations, as in the case of the $2$-element POVM, on the term in the target qubit state, corresponding to the second output of the $(i-1)^{th}$ module.
To do this, consider the $(i-1)^{th}$ and the $i^{th}$ modules of a POVM module sequence, and the ancilla register states corresponding to their pairs of outcomes, $\{|o^{(i-1)}_1\rangle,|o^{(i-1)}_2\rangle\}$ and
$\{|o^{(i)}_1\rangle,|o^{(i)}_2\rangle\}$ respectively .
The explicit steps for constructing the $i^{th}$ module are:

\begin{enumerate}[label=(\alph*)]
\item Entangle the first POVM outcome of the $i^{th}$ module with the ancilla register state used for the second outcome of the $(i-1)^{th}$ module, which is "redirected" to the $i^{th}$ module so it can be "reused". Hence we get $|o^{(i)}_1\rangle=|o^{(i-1)}_2\rangle$.
\item Entangle the second outcome of the $i^{th}$ module with the free computational ancilla register state with smallest binary value such that it differs by just one qubit (additional $1$ in its binary expression) from $|o^{(i)}_1\rangle$.
If there are no free ancilla register states, add another ancilla qubit in initial state $|0\rangle$.
\item A $\theta^{(i)}_1$ and a $\theta^{(i)}_2$  $y$-rotation (similar to eq. (\ref{eq:theta12})), are performed on the ancilla qubit, differing between the $|o^{(i)}_1\rangle$ and $|o^{(i)}_2\rangle$ states. These two rotations are controlled by the other ancilla qubits, having the same values as in $|o^{(i)}_i\rangle$, and the target qubit in state $|0\rangle$ and $|1\rangle$ respectively.
\item The ancilla state entangled to the second output of $i^{th}$ module is changed to the unused ancilla register state with smallest binary value. This can be done by applying at most $(\ceil{\log_2i}-1)$ multi-qubit controlled-NOT gates. This is not a necessary step, but it ensures that all POVM outcomes are entangled to ancilla register states in order of increasing binary value.

\item Finally, $\hat{V}^{(i)}_1$ and $\hat{V}^{(i)}_2$ general unitary operations, are performed on the terms of the target qubit state corresponding to the two POVM outcomes of the $i^{th}$ module, entangled to $|o_1^{(i)}\rangle$ and $|o_2^{(i)}\rangle$ respectively. Each of these two unitaries is performed by two $\ceil{\log_2i}$-qubit-controlled-rotation gates.

\end{enumerate}
Following these steps the $i^{th}$ module transforms the joint state of the target and the ancilla systems, after the $(i-1)^{th}$ module as

\begin{multline}
|\Psi_{i-1}\rangle = \sum^{i-2}_{k=0} \big[
\hat{M}_{k+1}|\psi_0\rangle\big]|k\rangle +  \Bigg(\prod_{k=1}^{i-1}
\hat{V}_2^{(k)}
\hat{D}^{(k)}_2  \Bigg)
\hat{U}|\psi_0\rangle|i-1\rangle \rightarrow  \\ |\Psi_i\rangle = \sum^{i-2}_{k=0} \big[
\hat{M}_{k+1}|\psi_0\rangle \big] |k\rangle +
\hat{V}^{(i)}_1
\hat{D}^{(i)}_1  \Bigg(\prod_{k=1}^{i-1}
\hat{V}_2^{(k)}
\hat{D}^{(k)}_2  \Bigg)
\hat{U}|\psi_0\rangle |i-1\rangle +  \Bigg(\prod_{k=1}^{i}
\hat{V}_2^{(k)}
\hat{D}^{(k)}_2  \Bigg)
\hat{U}|\psi_0\rangle|i\rangle \ \ \ \ \ \ \ \ \ \ \ \ \ \ \ \ \ \ \ \\ =\sum^{i-1}_{k=0} \big[
\hat{M}_{k+1}|\psi_0\rangle \big] |k\rangle +  \Bigg(\prod_{k=1}^{i}
\hat{V}_2^{(k)}
\hat{D}^{(k)}_2  \Bigg)
\hat{U}|\psi_0\rangle|i\rangle, \ \ \ \ \ \ \ \ \ \ \ \ \ \ \ \ \ \ \ \ \ \ \ \ \ \ \ \ \ \ \ \ \ \ \ \ \ \ \ \ \ \ \ \ \ \ \ \ \ \ \ \ \ \ \ \ \ \ \ \ \ \ \ \ \ \ \ \
\end{multline}
where $
\hat{D}_1^{(k)} = \cos \theta_1^{(k)} |0\rangle \langle 0| + \cos \theta_2^{(k)} |1\rangle \langle 1|$ and $
\hat{D}_2^{(k)} = \sin \theta_1^{(k)} |0\rangle \langle 0| + \sin \theta_2^{(k)} |1\rangle \langle 1|$.
In this way we can obtain the output state in eq. (\ref{eq:nPOVM_out}) with measurement operators given by eq. (\ref{eq:Mi}).
Additionally the ancilla register state entagled to the $k^{th}$ outcome of the POVM is $|k-1\rangle$, the state with binary value $(k-1)$.
Constructing an iterative program, which performs the same steps for each module is straightforward.
The example of constructing a $3$-element POVM is included in appendix \ref{app:3POVM}.

\bigskip

\begin{figure}[H]
\centering

\includegraphics[width=1.0\textwidth]{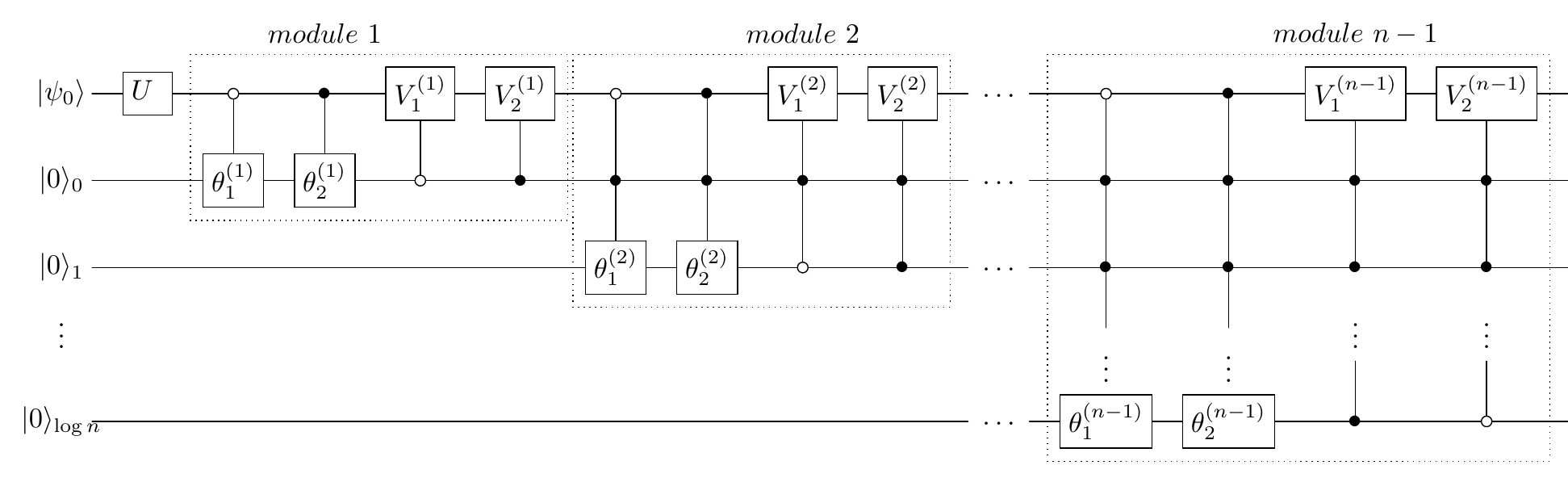}
    \caption{Representation of a quantum circuit performing an $n$-element POVM as a sequence of $(n-1)$ $2$-element POVM modules. The target qubit is in initial staet $|\psi_0\rangle$ and each of the $\ceil{\log_2 n}$ ancilla qubits is in initial state $|0\rangle$. Each module consists of two controlled $y$-rotation gates, and two controlled general unitary gates. For the $k^{th}$ module, these operations are controlled by the state of $\ceil{\log_2 k}$ qubits. The circuit has a maximum depth of $O(n^2)$ in terms of CNOT gates.\\
    $\_\_\_\_\_\_\_\_\_\_\_\_\_\_\_\_\_\_\_\_\_\_\_\_\_\_\_\_\_\_\_\_\_\_\_\_\_\_\_\_\_\_\_\_\_\_\_\_\_\_\_\_\_\_\_\_\_\_\_\_\_\_\_\_\_\_\_\_\_\_\_\_\_\_\_\_\_\_\_\_\_\_\_\_\_\_\_\_\_\_\_\_\_\_\_\_\_\_\_\_\_\_\_\_\_\_\_\_\_\_\_\_\_\_\_\_\_\_\_\_\_\_\_\_\_\_\_\_\_\_\_\_\_\_\_\_\_\_\_\_\_\_$}
    \label{fig:nPOVM_new}
\end{figure}

\section{Multi-qubit controlled operations and analysis of the complexity}\label{app:complexity}

Multi qubit controlled operations are used extensively in our POVM protocol.
To carry out a rotation around a single axis of the Bloch sphere of a qubit $q_0$, controlled by qubits $q_1..q_m$, the rotation is decomposed to two rotations with $m-1$ control qubits as
\begin{equation}
CR_i(\theta, q_1..q_m, q_0) =  CNOT(q_1, q_0)CR_i(-\frac{\theta}{2}, q_2..q_m, q_0)CNOT(q_1, q_0)CR_i(\frac{\theta}{2}, q_2..q_m, q_0)
\end{equation}
where $i\in \{x,y,z\}$ and $CR$ stands for controlled-rotation. By decomposing each controlled rotation further, the overall operation can be brought down to $(2^m-2)$ CNOTs and $2m$ one-qubit rotations.
Therefore the complexity of this method is exponential with $m$ - the number of control qubits.
An alternative method, suggested in \cite{nCU, RJway}, has linear complexity in terms of $m$, however it needs $m-1$ additional ancilla qubits.
For the examples of a $2$- and $3$-element POVMs considered in this paper the exponetial method is preferred, which, for the case of two-qubit controlled gates, has the same complexity and circuit depth as the linear method (both require $6$ CNOTs and $2$ single-qubit rotations), but does not need an additional ancilla qubit.
Nevertheless when implementing many-element POVMs, the use of the linear method should be considered.

To find the overall complexity of the protocol for a $n$-element POVM in terms of number of quantum gates, consider first the complexity of a single module. The  $i^{th}$ module requires up to $6$  $\ceil{\log_2 i}$-qubit controlled operations. Therefore its complexity is either $O(i)$ or $O(\log i)$ respectively, depending if the exponential or the linear method for a multi-qubit controlled operations is used.
The depth of the circuit for the $i^{th}$ module in these two cases is linear with $i$ - $0(i)$, or constant - $O(1)$ respectively.

\section{Constructing a circuit for the second POVM module}\label{app:3POVM}

This section illustrates the procedure for constructing the $i^{th}$ POVM module with the explicit example of the second module of a $3$-element POVM. The $3$ POVM outcomes require a $3$ dimensional ancilla space, therefore we need $2$-qubit ancilla register. Starting with the output state of the first POVM module, the system state can be written as

\begin{equation}
|\Psi \rangle = \Big(
\hat{V}_1^{(1)}
\hat{D}_1^{(1)}
\hat{U}|\psi_0\rangle\Big)|00\rangle + \Big(
\hat{V}_2^{(1)}
\hat{D}_2^{(1)}
\hat{U}|\Psi_0\rangle\Big)|10\rangle =
\hat{M}_1|\psi_0\rangle|00\rangle + (c|0\rangle+d|1\rangle)|10\rangle
\end{equation}
where an additional ancilla qubit in state $ |0\rangle $ is added, and $c$ and $d$ are coefficients such that $(c|0\rangle+d|1\rangle)=
\hat{V}_2^{(1)}
\hat{D}_2^{(1)}U|\psi_0\rangle$. Now we carry out the steps outlined in appendix \ref{app:nPOVM_steps}:

\begin{enumerate}[label=(\alph*)]

\item Associate the two outcomes of the $2^{nd}$ module, with ancilla register states $|o^{(2)}_1\rangle =|10\rangle$ and $|o^{(2)}_2\rangle =|11\rangle$ (at the end we will change $|o_2^{(2)}\rangle$ to $|01\rangle$).

\item Perform a $\theta_1^{(2)}$ and a $\theta_2^{(2)}$ y-rotations over the second ancilla qubit controlled by the first ancilla qubit in state $|1\rangle$, and the target qubit in states $|0\rangle$ and $|1\rangle$ respectively.

\begin{equation}
|\Psi\rangle \rightarrow |\psi_1\rangle|00\rangle+ \big(c\cos\theta_1^{(2)}|0\rangle + d\cos\theta_2^{(2)}|1\rangle\big)|10\rangle + \big(c\sin\theta_1^{(2)}|0\rangle + d\sin\theta_2^{(2)}|1\rangle\big)|11\rangle
\end{equation}

\item Using a doubly-controlled $X$ gate (equivalent to Toffoli gate) change $|o^{(2)}_2\rangle \rightarrow |01\rangle$, so that the POVM outcomes are entangled to states ordered in increasing binary value (taking the leftmost qubit as the least significant bit). Hence

\begin{equation}
|\Psi\rangle \rightarrow  |\psi_1\rangle|00\rangle+ \Big(
\hat{D}_1^{(2)}
\hat{V}_2^{(1)}
\hat{D}_2^{(1)}
\hat{U}|\Psi_0\rangle\Big)|10\rangle +  \Big(
\hat{D}_2^{(2)}
\hat{V}_2^{(1)}
\hat{D}_2^{(1)}
\hat{U}|\psi_0\rangle\Big)|01\rangle
\end{equation}.

\item Unitary operations $
\hat{V}_1^{(2)}$ and $
\hat{V}_2^{(2)}$, are performed on the target qubit, controlled by the ancilla register states $|o_1^{(2)}\rangle$ and $|o_2^{(2)}\rangle$ respectively. The system state at this point can be expressed as
\begin{equation}\label{3POVM_general}
|\Psi\rangle = (
\hat{M}_1|\psi_0\rangle)|00\rangle + (
\hat{M}_2|\psi_0\rangle)|10\rangle + (
\hat{M}_3|\psi_0\rangle)|01\rangle,
\end{equation}
where
\begin{equation}
\hat{M}_1 = \hat{V}_1^{(1)}\hat{D}_1^{(1)}\hat{U}
\end{equation}

\begin{equation}
\hat{M}_2 = \hat{V}_1^{(2)}\hat{D}_1^{(2)}\hat{V}_2^{(1)}\hat{D}_2^{(1)}\hat{U}
\end{equation}

\begin{equation}
\hat{M}_2 = \hat{V}_2^{(2)}\hat{D}_2^{(2)}\hat{V}_2^{(1)}\hat{D}_2^{(1)}\hat{U}
\end{equation}

\end{enumerate}

\end{document}